\documentclass[%
reprint,
superscriptaddress,
 amsmath,amssymb,
 aps,
 prl,
]{revtex4-1}

\usepackage[dvipsnames]{xcolor}
\definecolor{mblue}{RGB}{0, 114, 188}
\usepackage{hyperref}
\hypersetup{backref,pdfpagemode=FullScreen,colorlinks=true,breaklinks,urlcolor=mblue,linkcolor=mblue,citecolor=mblue}
\usepackage{graphicx}
\usepackage{dcolumn}
\usepackage{bm}
\usepackage{physics}
\usepackage{braket}
\usepackage{amssymb}

\begin{document}

\title{Universal relations between atomic dipolar relaxation and van der Waals interaction}

\author{Yuan-Gang Deng}
\thanks{These authors contributed equally to this work.}
\affiliation{State Key Laboratory of Low Dimensional Quantum Physics, Department of Physics, Tsinghua University, Beijing 100084, China.}
\affiliation{Guangdong Provincial Key Laboratory of Quantum Metrology and Sensing $\&$ School of Physics and Astronomy, Sun Yat-Sen University (Zhuhai Campus), Zhuhai 519082, China.}

\author{Yi-Quan Zou}
\thanks{These authors contributed equally to this work.}
\affiliation{State Key Laboratory of Low Dimensional Quantum Physics, Department of Physics, Tsinghua University, Beijing 100084, China.}

\author{Gao-Ren Wang}
\thanks{These authors contributed equally to this work.}
\affiliation{School of Physics, Dalian University of Technology, Dalian 116024, China.}

\author{Qi Liu}
\affiliation{State Key Laboratory of Low Dimensional Quantum Physics, Department of Physics, Tsinghua University, Beijing 100084, China.}

\author{Su Yi}
\email{syi@itp.ac.cn}
\affiliation{CAS Key Laboratory of Theoretical Physics, Institute of Theoretical Physics, Chinese Academy of Sciences, P.O. Box 2735, Beijing 100190, China.}

\author{Meng Khoon Tey}
\email{mengkhoon\_tey@tsinghua.edu.cn}
\affiliation{State Key Laboratory of Low Dimensional Quantum Physics, Department of Physics, Tsinghua University, Beijing 100084, China.}
\affiliation{Frontier Science Center for Quantum Information, Beijing, China.}

\author{Li You}
\email{lyou@tsinghua.edu.cn}
\affiliation{State Key Laboratory of Low Dimensional Quantum Physics, Department of Physics, Tsinghua University, Beijing 100084, China.}
\affiliation{Frontier Science Center for Quantum Information, Beijing, China.}

\date{\today}

\begin{abstract}
Dipolar relaxation happens when one or both colliding atoms flip their spins exothermically inside a magnetic ($B$) field. This work reports precise measurements of dipolar relaxation in a Bose-Einstein condensate of ground state $^{87}$Rb atoms together with in-depth theoretical investigations. Previous perturbative treatments fail to explain our observations except at very small $B$-fields. By employing quantum defect theory based on analytic solutions of asymptotic van der Waals interaction $-C_6/R^6$ ($R$ being interatomic spacing), we significantly expand the applicable range of perturbative treatment. We find the $B$-dependent dipolar relaxation lineshapes are largely universal, determined by the coefficient $C_6$ and the associated $s$-wave scattering lengths $a_{\rm sc}$ of the states before and after spin flips. This universality, which applies generally to other atomic species as well, implicates potential controls of dipolar relaxation and related cold chemical reactions by tuning $a_{\rm sc}$.
\end{abstract}

\maketitle

While nominally weak, the paradigm magnetic dipole-dipole interaction (MDDI) plays an essential role
in many important phenomena and systems, ranging from quantum many-body phases \cite{Lahaye09,Baranov12,Baier2016ExtendedHubbard,Yi07,Bottcher19,Chomaz19,Tanzi19,Schmitt16,Chomaz16}, to $d$-wave Feshbach resonances \cite{Cui17,Yao19,Boesten97} in atomic quantum gases, to diverse functional materials \cite{PhysRevLett.79.4669, PhysRevLett.87.047205} as well as protein folding~\cite{Dill1990} in biological systems. The very dipolar interaction also gives rise to inelastic relaxation inside an
external magnetic ($B$-) field, which together with three-body inelastic decay constitute the two leading mechanisms
limiting coherence times of ultracold atomic gases \cite{Hanson352,Dalibard98,Gerton99,Ketterle03,Hensler2003,Pasquiou10,Burdick15}, hence their applications in quantum science and technology. It is therefore essential to develop a general and clear understanding of how dipolar relaxation is affected by interatomic interactions in order to offer ways to control and suppress such relaxations.

\begin{figure}
\centering\includegraphics[width=1\columnwidth]{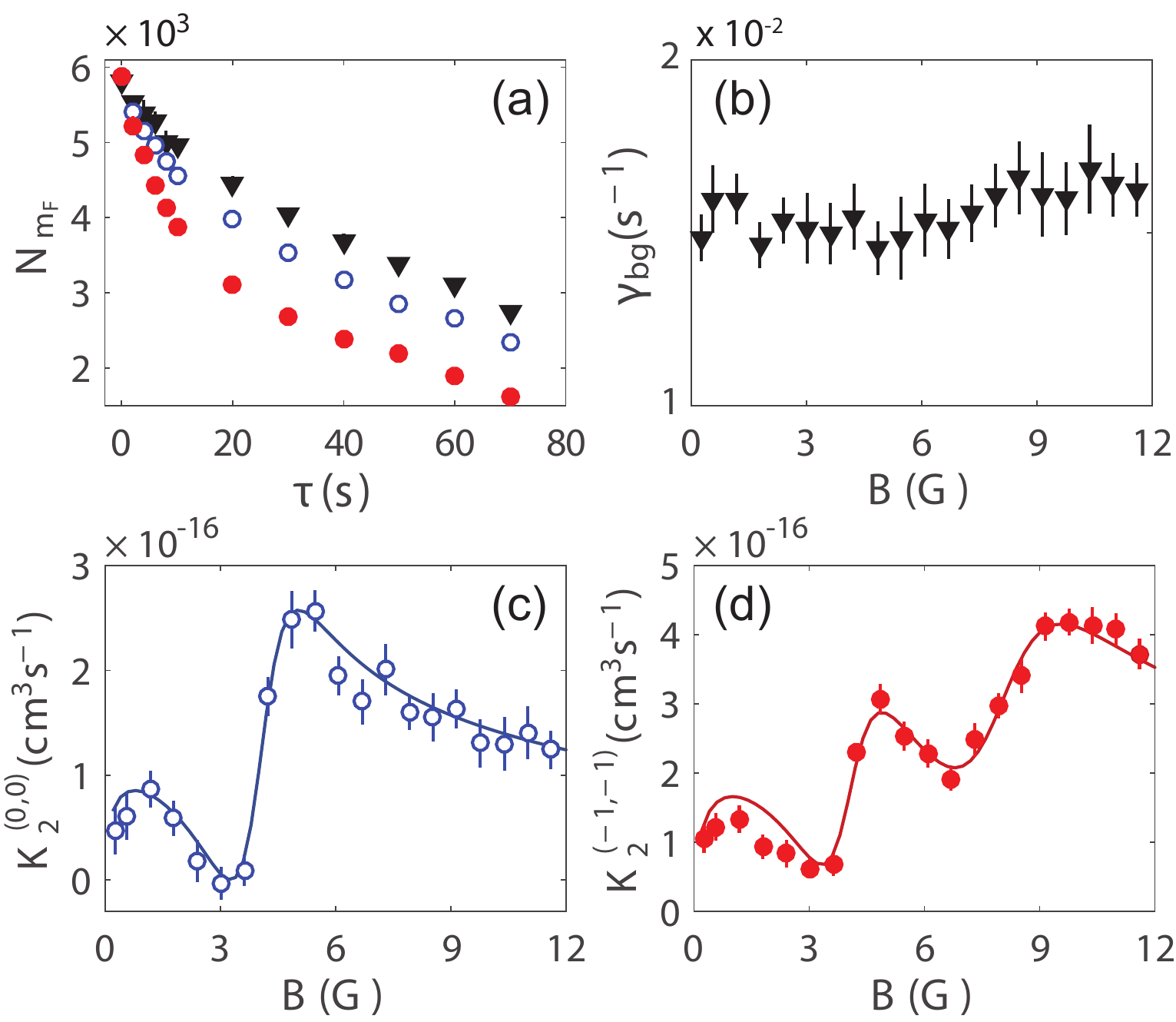}
\caption{Measured dipolar loss rates of $^{87}$Rb atoms in $F=1$. (a) The measured loss of BEC atoms prepared in $m_F=1$ (black triangles), $0$ (blue open circles), and $-1$ (red filled circles) states at $B=10.38$ G. (b) The measured loss rate (for $\tau<4$ s) of the lowest energy state ($m_F=1$) is essentially independent of magnetic field within the range we investigate. (c) and (d) display the measured two-body dipolar relaxation rates $K^{(m_F,m_F)}_2$ for atoms in $m_F=0$ or -1, respectively. Solid lines denote CC results using the state-of-the-art $^{87}$Rb potential~\cite{Strauss10}.
Experimental data are limited to $B<12$ G by allowed current to the field generating coils. Error bars represent one standard deviation over five measurements.}
\label{fig1}
\end{figure}

At zero magnetic field ($B=0$) and away from any resonance, cold elastic scattering cross sections from MDDI
are essentially energy independent and proportional to the fourth power of dipole moments \cite{you98,Hensler2003}. At finite field $B\neq 0$, two-body dipolar relaxation arises due to exothermic spin flips. Quantitative comparisons between measured dipolar losses \cite{Dalibard98,Gerton99,Hensler2003,Ketterle03,Pasquiou10,Burdick15} and theories remain limited, partly because effects due to MDDI are rather weak to measure accurately except in atoms with large magnetic dipole moments such as Cr \cite{Hensler2003,Pasquiou10} and Dy \cite{Burdick15}. Experimental observations
at small $B$-fields are largely consistent with theories based on Born approximation (BA) \cite{Hensler2003,Burdick15} which neglects the wave function dependence on interatomic interactions, or distorted-wave Born approximation (DBA) which partially accounts for phase shift of long-range wave function \cite{Pasquiou10}. Further improved understanding calls for more quantitative studies contrasting experiment and theory at larger $B$-fields, using accurate coupled-channel (CC) \cite{Tiesinga93,Mies96,Bohn09} calculations as reliable benchmark checks.

In this work, we measure $B$-field dependence of weak dipolar losses in a Bose-Einstein condensate (BEC) of $^{87}$Rb atoms with high accuracy. Different from earlier results in other atomic species~\cite{Hanson352,Dalibard98,Gerton99,Ketterle03,Hensler2003,Pasquiou10,Burdick15}, which include a small $B$ rise of $\propto \sqrt{B}$, we observe
a tail-off at large $B$ preceded by wavy structures. Upon detailed analysis of the measurement data, a simple scaling relation between one- and two-spin-flip dipolar-loss rates is revealed,
implicating the existence of a universal relationship underlying the two processes.
This is intriguing since atomic dipolar relaxations in heavier atoms
such as $^{87}$Rb are known to be strongly affected by short-range second-order spin-orbit interaction (SOI) \cite{Mies96} which operates on multichannel short-range wave functions not known to obey a simple relation. Extending earlier perturbative approaches \cite{Hensler2003,Burdick15,Pasquiou10}, we show that dipolar relaxation lineshapes from including both long-range MDDI and short-range SOI behave universally, largely determined by the interatomic van der Waals (vdW) interaction $-C_6/R^6$ plus the respective $s$-wave scattering lengths $a_{\rm sc}$ of the initial and final spin-flipped channels.

{\em Experiment} --- Our experiments are carried out in a BEC
of $F=1$ $^{87}$Rb atoms confined by an optical dipole trap (see supplemental material~\cite{SM} for details). The high atom number detection resolution of our setup (calibrated using quantum shot noise of coherent spin state) \cite{Luo17,Zou18}, together with small fluctuations of condensate atom numbers,
enables small spin-flip losses (3 to 4 orders of magnitude smaller than those of Cr or Dy of the same densities) to be accurately measured.
Figure \ref{fig1}(a) shows the remaining atom numbers (as functions of holding time) for a pure BEC prepared in single $m_F$ ($=-1,0,1$) states at $B=10.38$\,G.
The lowest energy $m_F=1$ state, which experiences no two-body dipolar loss, shows the longest $1/e$ lifetime and serves to calibrate an essentially $B$-independent background loss rate of $\gamma_{\rm{bg}} \simeq 0.0159 \pm 0.0005$ $\mathrm{s^{-1}}$ for all spin states [Fig.~\ref{fig1}(b)].
The two-body loss coefficients $K_2^{(0,0)}$ and $K_2^{(-1,-1)}$ are extracted for atoms initially prepared in $m_F=0$ [Fig.~\ref{fig1}(c)] or $-1$ state [Fig.~\ref{fig1}(d)], respectively~\cite{SM}.
Their entirety, including the wavy behaviors sandwiched in between the initial rise and tail-off,
are essentially reproduced by CC calculations (solid lines) without any fitting parameters, which corroborates the excellent accuracy of the full $^{87}$Rb molecular potentials \cite{Strauss10} with the quality of experimental data.

\begin{figure}
\centering\includegraphics[width=1\columnwidth]{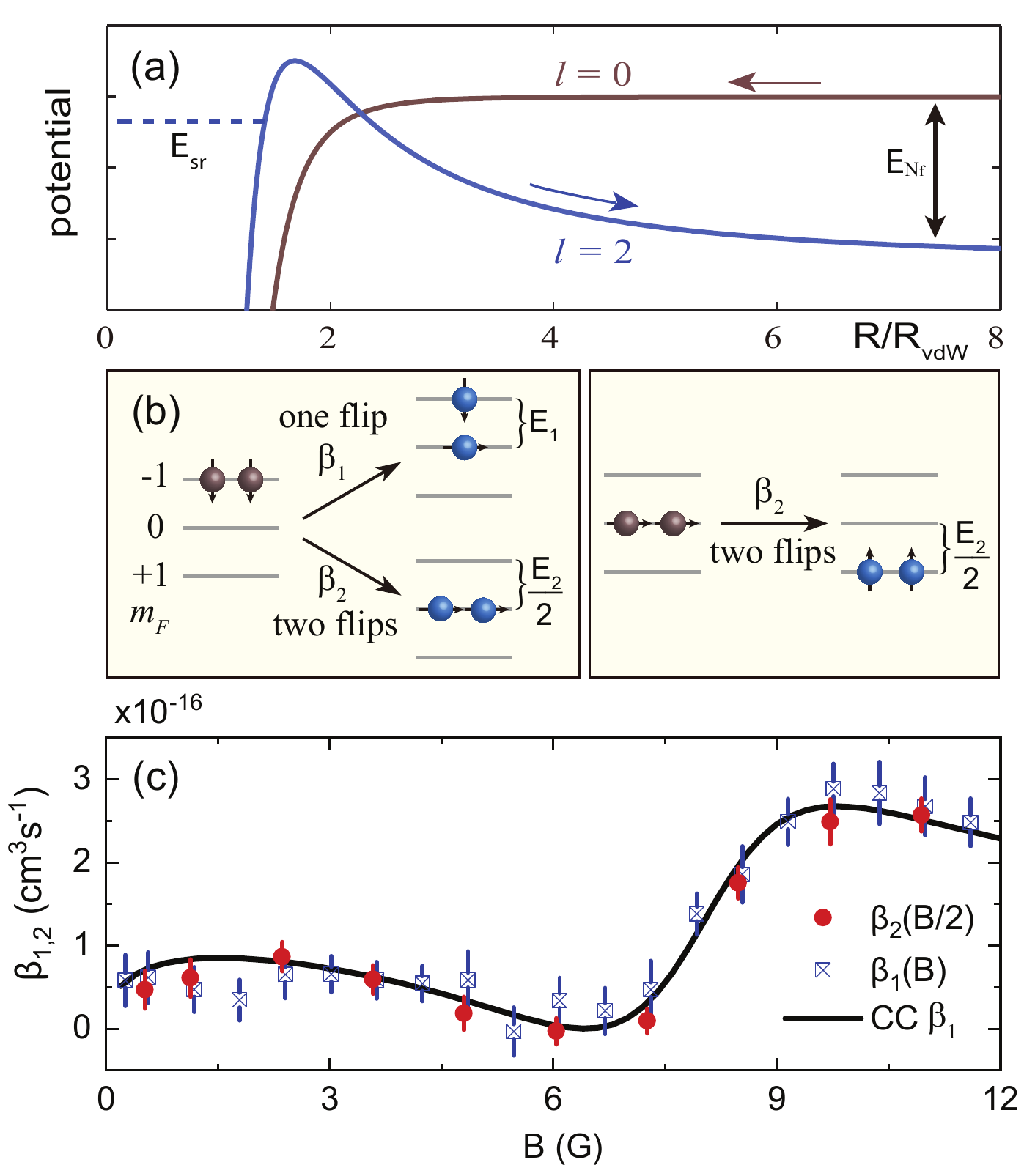}
\protect\caption{Self-similarities in dipolar loss lineshapes. (a) Dipolar relaxation occurs by flipping atomic spin, transforming spin angular momentum into relative atomic motion. At condensate temperatures, the incoming channel is $s$-wave while the spin-flipped outgoing channel is $d$-wave. The latter features a centrifugal barrier having great influence on the dipolar loss rates. For $^{87}$Rb, a $d$-wave shape resonance exists at $E_\mathrm{sr}= 1.04 E_{\rm{vdw}}$ (dashed line) above the $d$-wave threshold, where $E_\mathrm{vdW}$ is the vdW energy scale \cite{vdWscale}. (b) The allowed dipolar loss channels for atoms prepared in $m_F=-1$ or 0. As the energy $E_{N_f}$ released through $N_{f}$ spin flips is normally much larger than trap depth, both atoms are lost after spin flips. (c) The extracted one-flip loss rate [$\beta_1(B)=K^{(-1,-1)}(B)-K^{(0,0)}(B)$] (crossed open squares) and two-flip rate data with scaled $B$-field [$\beta_2(B/2)=K^{(0,0)}(B/2)$] (red solid circles). The solid line denotes CC results for $\beta_1$ with both MDDI and SOI included.
}
\label{fig2}
\end{figure}

{\em Perturbative model} --- The atomic spin-spin interaction responsible for dipolar relaxations is (in atomic units)
\begin{equation}
V_{\rm{ss}}({\bf R})=\alpha^2(\frac{1}{R^3}+\frac{\kappa_{\rm{so}}}{e^{b_{\rm{so}}R}})[{{\bf S}_1\cdot {\bf S}_2 -3({\bf S}_1\cdot {\hat{{{R}}}})({\bf S}_2\cdot {\hat {{R}}})}]
\label{vss}
\end{equation}
for two atoms located at ${\bf r}_1$ and ${\bf r}_2$ \cite{Strauss10,Yi2000}.
It consists of MDDI $\propto 1/R^3$ and second-order SOI $\propto \kappa_{\rm{so}}\exp(-b_{\rm{so}}R)$
with identical electron spin dependence. Here, $\alpha$ is the fine structure constant, $R =|{\bf R}|=|{\bf r}_1 -{\bf r}_2|$, ${\hat {{R}}} = {\bf R}/R$. For $^{87}$Rb atoms, $\kappa_{\rm{so}}=-9.1829\, a_B^{-3}$ and $b_{\rm{so}}= 0.7196\,a_B^{-1}$ (Bohr radius $a_B$) \cite{Mies96,Strauss10}. The magnitude of SOI becomes larger than MDDI for $R<16.8\,a_B$, a separation much shorter than the vdW length scale \cite{vdWscale} ($=82.7$ $a_B$ for $^{87}$Rb atoms). 
At small $B$, dipolar relaxation selection rules can be discussed using the hyperfine spin $\mathbf{F=S+I}$ ($S=1/2$ and $I=3/2$ for $^{87}$Rb). Inside a $B$-field along $z$-axis, the $z$-component of the total angular momentum, $m+m_{F1}+m_{F2}$, remains conserved in the presence of $V_{\rm{ss}}$. Here, $m_{Fj}$ and $m$ are the eigenvalues of, respectively, $F_{jz}$ for atom $j=1,2$, and $l_z$, the $z$-component of molecular orbital angular momentum.
$m$ changes during inelastic dipolar relaxation at the expense of
$m_{Fj}$, constrained by the selection rule $\Delta l=0,\pm2$ \cite{SM}.
At BEC temperature, dipolar relaxation thus corresponds to scattering from
exclusively $s$-wave ($l=0$) to $d$-wave ($l=2$) channel \cite{Pasquiou10} through $V_{\rm{ss}}({\bf R})$ as illustrated in Fig. \ref{fig2}(a).

Adopting distorted-wave \cite{Messiah69} and single-channel \cite{singlechannel} approximation, we show that the total cross section for dipolar relaxation from channel $a$ to $b$ can be approximated by \cite{Fbasis,SM}
\begin{eqnarray}\label{eq:totalcrosssection}
\sigma_{a\rightarrow b}&\approx&\frac{1}{E_a^{3/2}E_b^{1/2}}\left|\int_0^\infty G_2^{b\ast}(R)\widetilde{U}_\mathrm{ss}(R)G_0^a(R)dR \right|^2 \nonumber\\
&&\times\left|\langle \varphi_b|({\bf F}_1\otimes {\bf  F}_2)^2_{N_f}|\varphi_a\rangle\right|^2.
\end{eqnarray}
Here, $G^{b(a)}_{l}(R)$ and $E_b$($E_a$) denote, respectively, the radial wave function and the kinetic energy of the outgoing(incoming) wave. $({\bf F}_1\otimes {\bf  F}_2)^2_{N_f}$ is the rank-2 spherical tensor with $N_f=1,2$ for one and two spin flips, respectively.
For atoms in $F=1$ ground states, direct spin-flip losses can occur only in a total of seven channels as dictated by the spin part in Eq.~(\ref{eq:totalcrosssection}). For atoms prepared in $m_F=-1$, the incident spin state
$|\varphi_a\rangle=|m_{F_1}=-1;m_{F_2}=-1\rangle$,
the corresponding one- and two-flip final spin states are
$|\varphi_b^{(N_f=1)}\rangle =\left(|-1;0\rangle +|0;-1\rangle\right)/{\sqrt{2}}$
and $|\varphi_b^{(N_f=2)}\rangle =|0;0\rangle$. Each spin flip changes $m_{Fj}$
by $1$ ($\hbar$) and a maximum of $2$ ($\hbar$) occurs for two flips.
In the process, an energy of $E_{N_f}=|N_fg_F \mu_B B|$ \cite{Lande} is released, giving $E_b=E_a+E_{N_f}$.

The nearly identical $s$-wave scattering lengths of $F=1$ $^{87}$Rb:
$a_0=101.8\,a_B$ ($a_2=100.4\,a_B$) for total spin (${\bf F}_1+{\bf F}_2$) of ``0" (``2"), results in an approximate SU(2) collision symmetry, which encourages treating scattering approximately by a single channel (apart from the spin and orbital angular parts) with $a_{\rm sc}=a_2\approx a_0$ for arbitrary spin states. Based on Eq.~(\ref{eq:totalcrosssection}), the dipolar losses for the seven channels are thus describable by
two irreducible rates $\beta_1$ and $\beta_2$, respectively for one- and two-spin flips \cite{SM}. Dipolar relaxation for atoms prepared in $m_F=-1$ includes contributions from both one- and two-spin flips, i.e. $K_2^{(-1,-1)}=\beta_1+\beta_2$, while only two-flip process is allowed for atoms in $m_F=0$, i.e. $K_2^{(0,0)} =\beta_2$ [Fig. \ref{fig2}(b)].
The one-flip rate $\beta_1$ can thus be determined using $\beta_1=K^{(-1,-1)}_2-K^{(0,0)}_2$ based on experimental data, and is found to be remarkably self-similar to $\beta_2$ according to $\beta_1(B)\approx\beta_2(B/2)$ [Fig. \ref{fig2}(c)].

To illuminate the physics behind the observed lineshapes
including their self-similar dependence on $B$, we expand
earlier perturbative approaches to larger $B$-field range (beyond BA \cite{Hensler2003,Burdick15} and DBA \cite{Pasquiou10}) using a single-channel semi-analytic quantum defect theory (QDT) \cite{singlechannel}, assuming that both incoming and outgoing wave functions are distorted by the same vdW interaction down to small $R$ \cite{Gao01}. As QDT using vdW potential ignores short-range details of the wave functions, which subsequently lead to inaccuracies in treating the SOI term, we will first assume $\widetilde{U}_{ss}=\alpha^2 (g_F^2/g_s^2)\sqrt{24\pi/5}/R^3$ \cite{Lande}, i.e., treating only the MDDI term (see \cite{SM} for relation between $\widetilde{U}_{ss}$ and $V_{\rm{ss}}(\mathbf{R})$). The loss rate per unit particle density is calculated according to $\beta_{a\rightarrow b} =  \langle\sigma_{a\rightarrow b} v_a\rangle_a$ by averaging over the distribution of the incoming velocity $v_a=(2E_a/\mu)^{1/2}$ of the gas ensemble in the initial spin state. The appropriate $G^{q=a,b}_l(R)$ is normalized to $\sin(k_qR-{l\pi}/{2}+\delta^q_l)$ at large $R$ with $\delta^q_l$ being the phase shift for the $l$-th partial wave.

The differences between BA \cite{Hensler2003,Burdick15}, DBA \cite{Pasquiou10}, and our approach can be highlighted using $G^{a,b}_{l}(R)$. BA completely ignores the effects of interatomic potential
on the scattering wave functions, giving $G_0^a(R)=k_a R\,j_0(k_a R)$ and $G_2^b(R)=k_b R\,j_2(k_b R)$ \cite{Bessel} which correspond, respectively, to the $s$- and $d$-partial wave components of the associated (incoming or outgoing) plane waves. The DBA model adopted in Ref.~\cite{Pasquiou10} inserts a phase shift $\delta_0^a$ into the long-range wave function of the incoming $s$-wave, but ignores any distortion to the outgoing $d$-wave, leading to $G_0^a={(k_a R)(1+e^{i2\delta_0^a})}\left[j_0(k_aR)+k_aa_{\rm sc} y_0(k_aR)\right]/{2}$ \cite{Bessel} with $-\tan\delta_0^a=k_a a_{\rm sc}^a$. Our approach accounts for the effects of the vdW potential on both the incoming and outgoing wave functions, by using the semi-analytic solutions of $-C_6/R^6$ for all $R$ and replacing the effect of short-range potential by a single (quantum defect) parameter uniquely related to $a_{\rm sc}$.
This is referred to as vdW universality for cold atom collision \cite{Gao01},
whereby $G_2^b(R)$ and $G_0^a(R)$ are determined by just three parameters,
namely, $C_6$, $E_{a,b}$, and $a_{\rm sc}^{a,b}$.
With our approach, such a universality clearly permeates to dipolar relaxation
given by Eq. (\ref{eq:totalcrosssection}) as well. The overall $B$-dependent
lineshapes for one or two spin-flips thus take the same functional form
depending only on $C_6$ and $a_{\rm sc}^{a,b}$.

\begin{figure}
\centering\includegraphics[width=0.95\columnwidth]{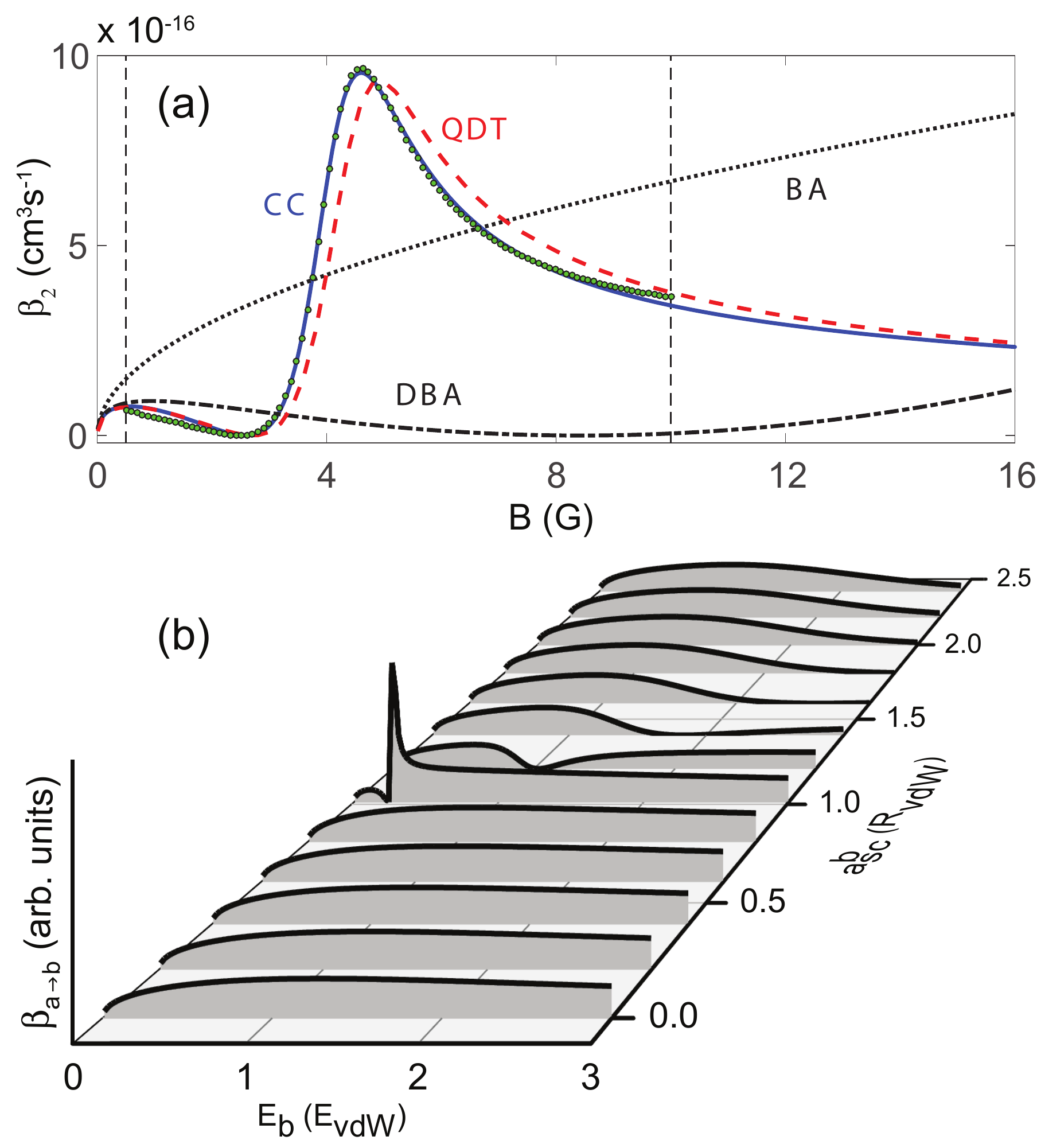}
\caption{Universal lineshape. (a) Calculated $\beta_2$ from MDDI (without SOI) using perturbative BA, DBA, QDT, and full CC. Green dots show a fit of the CC results using Fano lineshape with an asymmetry parameter $q=1.8$ for $0.5<B<10$ G \cite{SM}.  (b) Universal lineshapes for dipolar loss rate $\beta_{a\rightarrow b}$ as a function of $E_b$ and $a_\mathrm{sc}^b$ of the spin flipped state, both in vdW scales. $a_\mathrm{sc}^a/R_\mathrm{vdW}=0.5$ for the initial spin state in this plot. For ultracold collisions, $\beta_{a\rightarrow b}$ is essentially independent of the incident energy $E_a$ \cite{SM}.
}
\label{fig3}
\end{figure}

{\em Universal lineshape} --- The computed $\beta_2$ based on the aforementioned perturbative models and CC calculations are compared in Fig.~\ref{fig3}(a) (all with SOI ignored). For the demonstrated $B$-field range, BA and DBA fail largely to agree with the CC results, whereas the QDT model works uniformly well. For BA, the prediction $\beta_2\propto \sqrt{B}$ \cite{SM} agrees with CC only for $B<0.04$ G. The DBA model~\cite{Pasquiou10} predicts vanishing dipolar loss approximately at $B_\vee=64\hbar^2/(9\pi^2|g_F|\mu_B \mu a^{a2}_\mathrm{sc})$
for positive $a_{\rm sc}^a$ \cite{SM} and expands the range of agreement with CC to about 0.2 G, but deviates seriously at higher $B$-fields. Our model successfully extends vdW universality to dipolar relaxation
 by explaining the observed self-similar patterns as well as the wavy structures.
Working out the constant prefactors associated with the spin part in Eq.~(\ref{eq:totalcrosssection}) and taking into account the factor of 2 in linear Zeeman shifts $E_{N_f}$ for one and two flips,
we find analytically the observed self-similar $B$-scaling relation $\beta_1(B)\approx\beta_2(B/2)$.

In more detail, we confirm the observed wavy structure arises as
a special case of vdW universality.
In a seminal paper, Gao predicts a universal $d$-wave quasi-bound state for vdW potential near threshold
when $a_{\rm sc}$ is slightly greater than 0.956$R_\mathrm{vdW}$. Our case of $^{87}$Rb ground state atoms corresponds to $a_{\rm sc}^b\approx a_{\rm sc}^a=a_{\rm sc}\approx1.21R_\mathrm{vdW}$, and the $d$-wave shape resonance
is shifted up slightly to $\sim 1.04 E_{\rm vdW}$ \cite{vdWscale}. Such $d$-wave shape resonances have been observed in $^{39}$K, $^{41}$K, and $^{174}$Yb \cite{Yao19,Burke99,Tojo06}, in $F=2$ $^{87}$Rb \cite{Boesten97,Thomas04,Buggle04}, and in $F=1$ $^{87}$Rb at a very high $B$ ($\sim632$ G) \cite{Volz05}. In dipolar relaxation, interference between this shape resonance with the continuum in the $d$-wave outgoing channel gives rise to the celebrated Fano lineshape \cite{fano} (as illustrated by the green dots in Fig.~\ref{fig3}(a)).
In fact, the observed wavy structures in Fig. \ref{fig1}(d) consist of two Fano profiles, respectively associated with one- and two-flip $d$-wave channels. Because this $d$-wave shape resonance lies close to the top of the centrifugal barrier, the observed resonance lineshapes are rather broad despite of being in a high partial wave.
More generally, for atomic species without SU(2) collision symmetry, we check that the vdW universality still holds. Specifically, for a finite range of $E_b$, there exists a unique relation between the dipolar relaxation rate $\beta_{a\rightarrow b}$ between the scaled spin-flipped energy $E_b/E_\mathrm{vdW}$ and the scaled $s$-wave scattering lengths $a_\mathrm{sc}^{a,b}/R_\mathrm{vdW}$, independent of small $E_a$. Figure \ref{fig3}(b) shows some examples of such universal lineshapes with $a_\mathrm{sc}^a/R_\mathrm{vdW}=0.5$, which highlights the effects of $d$-wave shape resonance near $a_\mathrm{sc}^{b}/R_\mathrm{vdW} \approx 1$.

\begin{figure}
\centering\includegraphics[width=0.95\columnwidth]{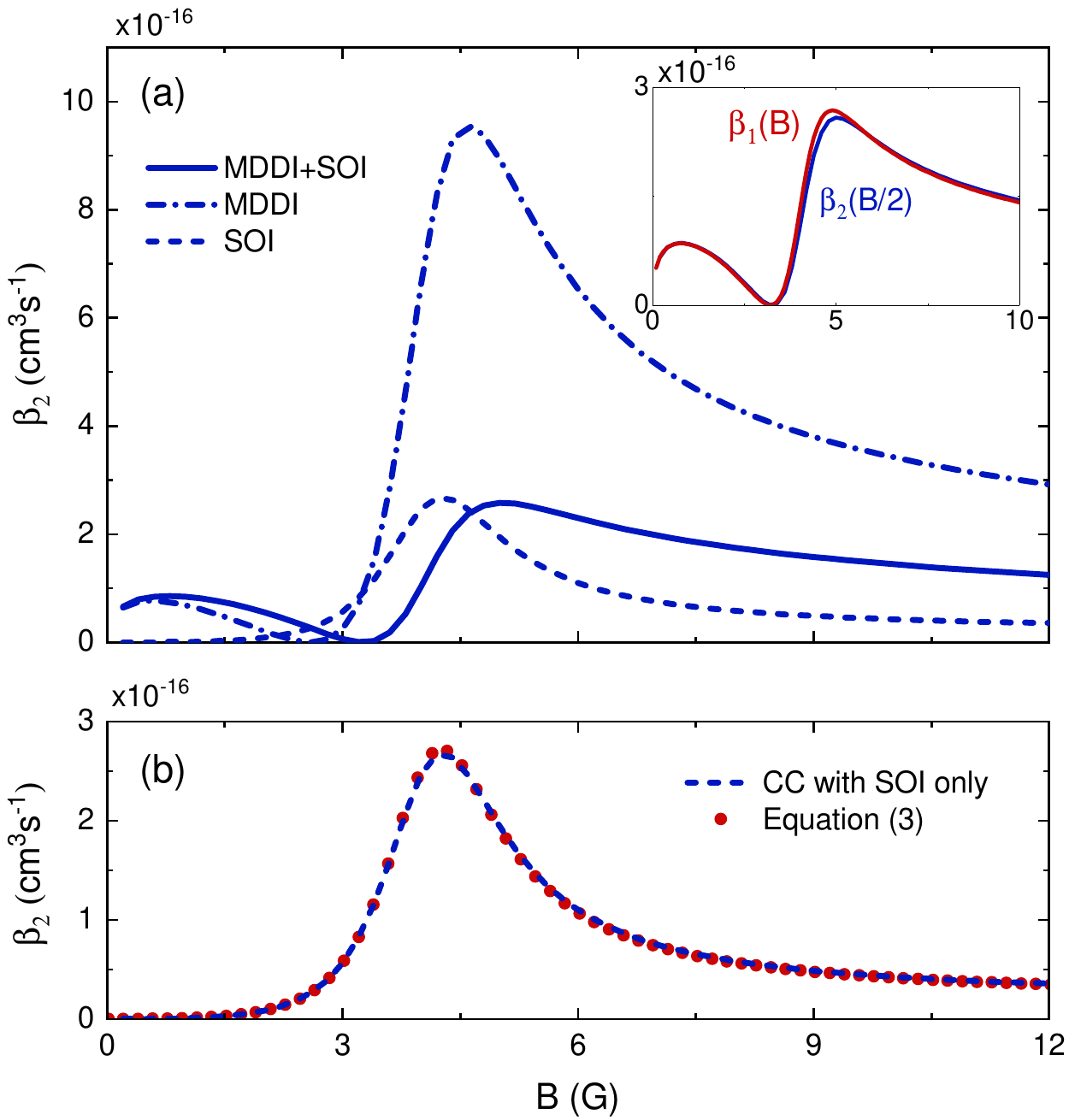}
\caption{Contributions of SOI and MDDI to dipolar loss. (a) The loss rate $\beta_2$ from CC as a function of $B$. The curve labeled ``MDDI'' (``SOI'') includes only MDDI (or second-order SOI) contribution, while ``MDDI+SOI'' denotes the inclusion of both.
The scaling relationship $\beta_1(B)\approx\beta_2(B/2)$ shown in inset survives even when SOI is included, suggesting that the dipolar loss due the short-range SOI is related to the long-range potential through a simple rule. (B) The CC results with only SOI agrees completely (up to a constant amplitude) with Eq.~(\ref{SOIlineshape}), which depends only on $B$, $a_{\rm sc}^b$, and $C_6$.}
\label{CCSO}
\end{figure}

The overall features of dipolar loss we observe are thus qualitatively explained by the effects of vdW potential using QDT with SOI ignored. We next compare in detail in Fig.~\ref{CCSO} the CC predictions using full $V_\mathrm{ss}$ (``MDDI+SOI'', solid line), MDDI (dashed-dotted line), or SOI (dashed line). For $B>2.6$~G, inclusion of SOI suppresses the dipolar relaxation (due to MDDI) considerably through interference effects.
Intriguingly, the scaling relation remains approximately the same when SOI is included (see inset of Fig.~\ref{CCSO}(a)). This is puzzling since SOI is more significant at small $R$ where the wave functions are of multichannel nature and differ for different outgoing channels. It turns out that the lineshape of SOI-induced dipolar loss is describable by an analytic function \cite{SM,SOIuniversality}
\begin{equation}\label{SOIlineshape}
\beta_\mathrm{SOI}\propto \left[(Z_{fs}^c-{\cal K}^cZ_{gs}^c)^2+(Z_{fc}^c-{\cal K}^cZ_{gc}^c)^2\right]^{-1},
\end{equation}
related solely to the vdW universality through $C_6$, $a_\mathrm{sc}^b$ and $E_b$. The validity of Eq.~(\ref{SOIlineshape}) is evident from its excellent agreement with the CC results including only SOI, up to a constant factor [Fig.~\ref{CCSO}(b)]. This understanding partially explains the scaling relationship for the SOI case, up to an unknown amplitude.

In summary, we show that atomic dipolar relaxation rates can be universally described by the long-range vdW potential plus the $s$-wave scattering lengths for the incoming and (spin flipped) outgoing channels within a finite spin-flipped energy. The distorted wave approximation we adopt based on QDT wave functions including the vdW potential greatly expands the applicability of perturbative prediction for dipolar loss, and is found to work well also for other alkali-metal atoms. In the future, it would be interesting to apply this model to other atomic species such as Dy, Cr, Er whose dipolar relaxations are orders of magnitude stronger, but detailed knowledge of their interatomic potentials are unavailable to support accurate CC predictions.

\begin{acknowledgments}
The authors thank Peng Zhang and Jinlun Li for helpful discussions. This work is supported by the National Key R\&D Program of China (Grant No. 2018YFA0306504, No. 2018YFA0306503 and 2018YFA0307500) and the NSFC (Grant No. 91636213, No. 11654001, No. 91736311 and 11874433).
\end{acknowledgments}

\bibliographystyle{apsrev4-1}

\begin{thebibliography}{45}
\bibitem [{vdW()}]{vdWscale}%
  \BibitemOpen
  \href@noop {} {}\bibinfo {note} {The vdW length and energy scales are defined
  by $R_\mathrm{vdW}=(2\mu C_6/\hbar^2)^{1/4}/2$ and
  $E_\mathrm{vdW}=\hbar^2/(2\mu R_\mathrm{vdW}^2)$, respectively. $\mu$ is the
  reduced mass.}\BibitemShut {Stop}%
\bibitem [{sin()}]{singlechannel}%
  \BibitemOpen
  \href@noop {} {}\bibinfo {note} {We assume that the incoming and outgoing
  scattering wave functions are of single channel nature, essentially ignoring
  the multi-channel nature of the short-range wave functions in real atoms.
  This approximation would be accurate if the contributions to dipolar
  relaxation come mainly from the long-range wave functions.}\BibitemShut
  {Stop}%
\bibitem [{Fba()}]{Fbasis}%
  \BibitemOpen
  \href@noop {} {}\bibinfo {note} {As our work concerns only dipolar relaxation
  within the same hyperfine manifold (since $F$ changing process is endothermic
  and thus forbidden), it is convenient to use the hyperfine spin $\mathbf{F}$
  operators, see \cite{SM} for formula in electron spin $\mathbf{S}$
  operator.}\BibitemShut {Stop}%
\bibitem [{Lan()}]{Lande}%
  \BibitemOpen
  \href@noop {} {}\bibinfo {note} {$g_F$ and $g_s$ are the Land\'{e} g-factors
  of the hyperfine state and of electron, respectively. $\mu_B$ is the Bohr
  magneton.}\BibitemShut {Stop}%
\bibitem [{SOI()}]{SOIuniversality}%
  \BibitemOpen
  \href@noop {} {}\bibinfo {note} {The $B$-dependence lineshape of SOI-induced
  dipolar loss can be derived by taking a cue from
  Eq.~(\ref{eq:totalcrosssection}). SOI only applies to short-range wave
  functions which are multichannel and not describable by our single channel
  QDT wave functions (which are more accurate in mid-to-long-range interatomic
  spacing, $R > 20$ $a_B$). Nevertheless, it is known that the form of the
  short-range wave function of an incoming (outgoing) channel is essentially
  independent of $E_a$($E_b$), since $E_a$($E_b$) is many orders of magnitude
  smaller than the interatomic potential at small $R$. As a result, the
  lineshape of the dipolar loss due to SOI should only depend on the amplitude
  variation of the short-range wave functions and on $E_{a,b}$ according to
  Eq.~(\ref{eq:totalcrosssection}). The amplitude and energy variations of the
  incoming $s$-wave are negligible. The variation of short-range amplitude of
  the outgoing $d$-wave can be obtained by propagating the unity-normalized
  long-range wave function to short-range using known analytic vdW solutions
  according to \cite{SM} \begin{equation} \beta_\mathrm{SOI}\propto
  \left[(Z_{fs}^c-{\cal K}^cZ_{gs}^c)^2+(Z_{fc}^c-{\cal
  K}^cZ_{gc}^c)^2\right]^{-1}. \end{equation} Here, $Z_{fs}^c$, $Z_{fc}^c$,
  $Z_{gs}^c$ and $Z_{gc}^c$ are analytic functions depending only on $l$, $E_b$
  and $C_6$ (ref.~\cite{Gao01}). They relate the short-range wave functions of
  a pair of vdW wave function basis to the long-range basis
  $\sin(k_sr_s-l\pi/2)$ and $\cos(k_sr_s-l\pi/2)$ \cite{SM}. The short-range
  K-matrix ${\cal K}^c$ is given by \begin{align} {\cal K}^c =
  \frac{[\Gamma(3/4) R_{\rm vdW}+\Gamma(5/4)a_{\rm sc}]\tan(\pi/8)}
  {\Gamma(5/4)a_{\rm sc} -\Gamma(3/4)R_{\rm vdW}}, \label{Kc} \end{align} with
  $\Gamma(.)$ the gamma function.}\BibitemShut {Stop}%
\end{thebibliography}

%

\end{document}